\documentclass[onecolumn,showpacs,preprintnumbers,amsmath,amssymb]{revtex4}

\usepackage{graphicx}
\usepackage{dcolumn}
\usepackage{bm}

\begin{document}

\newcommand{\half}{\mbox{$\frac{1}{2}$}}


\title{Bell's Theorem: A Critique  }
\author{Michael Clover}
\affiliation{%
Science Applications International Corporation\\
San Diego, CA \\
}

\date{\today}
\begin{abstract}
{By implicitly assuming that all possible Bell-measurements occur simultaneously, all proofs of Bell's Theorem violate Heisenberg's Uncertainty Principle.   This assumption  is made in the original form of Bell's inequality, in Wigner's probability inequalities, and in the ``nonlocality without inequalities''  arguments.
The introduction of time into derivations of these variants of Bell's theorem results in extra terms related to the time order of the measurements used in constructing correlation coefficients.  Since the same locality assumptions are made in the Heisenberg-compliant derivations of this paper, only time-independent classical local hidden variable theories are forbidden by violations of the original Bell inequalities;  time-dependent quantum local hidden variable  theories can  satisfy this new bound and agree with experiment.    We further point out that factorizable wavefunctions have been used to describe some EPR experiments and can be used to describe others. These will generate local de Broglie-Bohm trajectories in the description of the data.  This  second, independent,  line of argument also shows that violation of  Bell's inequality is only evidence that  Heisenberg's Uncertainty Principle cannot be ignored.
 }
\end{abstract}

\maketitle

\noindent

\section{Introduction}

With the paper of EPR~\cite{EPR35}, it seemed clear that   local realism implied the incompleteness of Quantum Mechanics.  A recent article by Whitaker~\cite{W04} suggests that an even stronger case can be made for incompleteness simply by noting that in the time after Alice measures a property of an entangled singlet decay and before Bob makes his measurement, an element of reality is known to exist for Bob -- an element that the Copenhagen Interpretation  says cannot be known until Bob's measurement is actually made.  Such arguments for incompleteness led to the creation of various hidden variable theories, among them the de Broglie-Bohm interpretation~\cite{B52}, also known as ``Bohmian mechanics''.  However, when this model was applied to the interpretation of singlet-decay coincidence experiments (EPR-B experiments), non-local forces were generated.  Bell~\cite{B71} developed a mathematical inequality, supposedly based only on ``local realism'', that set an upper bound of 2 on  certain experimental measurements.  Quantum mechanics predicted  $2\sqrt{2}$, which was taken to show the non-locality of Quantum mechanics, consistent with the de Broglie-Bohm analysis of such singlet states.

Since that time, every experimental  ``proof'' of non-locality has involved plugging experimental measurements into Bell's formula  and comparing  the numerical result to the bound in Bell's inequality.    Historically, Bell derived one expression for his inequality~\cite{B71}, and others ({\it e.g.}~\cite{CHSH69}) derived other forms, varying in the symmetry of the expression, but all sharing the same conclusion:  a ``local realistic'' theory (or a local Reality) should generate  results that satisfy the bound ({\it i.e.}$<2$).  In the last 35-40 years, experiments ({\it e.g.}~\cite{OU88, ADR82, WZ98, Rowe01}) have universally discovered that the bound is violated, which has been interpreted by some to mean that reality is non-local.

Proponents of local realism respond with two kinds of arguments.  The first argument is that experimentalists have not plugged -- and may never be able to fully plug -- certain (detector) efficiency loopholes~\cite{Sa04}, in which case  experimental values -- properly normalized -- have never actually violated Bell's inequality, although the experiment of Rowe, {\it et al.}~\cite{Rowe01} seems to answer this objection.  The second argument is that Bell's derivation relies on counterfactual reasoning~\cite{HP02a, HP02b, Ad00}, and therefore isn't valid.  Our critique will focus on the counterfactual issues.  

We begin by noting that an ambiguity frequently arises in various authors' works due to confusion between {\em elements}  of reality and {\em measurements} of those elements.  For example, an electron's spin, as an element of reality, is a vector of radius $\sqrt{3/4}\hbar$ pointing anywhere in a solid angle of  $4\pi$.  A measurement of that spin, however,  only gives the dichotomic values of $\pm \hbar/2$ for the component in the direction of the field.  The {\em measurement} is not the same as the {\em element of reality} for a realist.  We will point out that in various ``proofs'' of Bell's Theorem, the assumption of the simultaneous existence of ``elements of reality'' is equated to the simultaneous existence of the results of measurements of those elements, the latter being in direct contradiction to Heisenberg's Uncertainty Principle.  This conflation can result when an author, attempting to avoid the specification of any particular hidden variable model, claims that ``the measurement-results that weren't measured are the only hidden variables'' they will consider.  It should  not be surprising, then, that bounds derived from such a confusion are violated by Quantum Mechanics and experiment; it is surprising that this theoretical aspect of counterfactuals has not been identified before.

\subsection{Outline}

In the next section, we briefly recapitulate one of Bell's derivations of his inequality, pointing out the counterfactual assumption,  and then derive a ``factual'' version of the inequality by keeping track of the different times the individual  measurements are made.  Having done  this for Bell's inequality, we also reanalyze Wigner's derivation of a variant inequality, as well as some ``nonlocality without inequalities" derivations.  In all  cases, careful attention to  time introduces terms related to the difference of cross-correlations measured in different order, for example:
\begin{eqnarray}
S_{Bell} < 2 + |\langle A(t_<)A'(t_>) \rangle - \langle A'(t_<)A(t_>) \rangle| \ . \nonumber
\end{eqnarray}

The subsequent section proposes  an experiment -- that could be performed by pollsters interviewing married couples, asking ambiguous questions  -- that could measure such an effect in a macroscopic, non-quantum context. 

The penultimate section reexamines the same inequalities from the quantum mechanical standpoint, where the presence of operator commutator terms is shown to augment Bell's value of $2$. ({\it i.e.} $S_{Bell}^2 \le 4\pm \langle [\hat{A},\hat{A'}] [\hat{B'},\hat{B''}]\rangle$).  We then calculate that both an entangled wavefunction or certain products of unentangled  wavefunctions can both generate  Bell-violating terms, and point out one experiment that was successfully analyzed using just such a product-form  wavefunction.

We conclude by pointing out that while  de Broglie-Bohm hidden variable models will appear to be non-local if  non-factorizable ``post-selection'' entangled wavefunctions are used in the model, in all experiments to date a factorizable ``pre-selection'' wavefunction can also be used to generate a local model of the experimental correlations.   This in turn suggests that there is nothing in reality {\em or} quantum mechanics that cannot be treated in a completely local manner. 

\section{Deriving Bell's Theorem}

Bell's derivation  of the Bell/CHSH inequality~\cite{B71} starts by calculating the difference of two (theoretical) averages or correlation coefficients,
\begin{eqnarray}
\langle AB\rangle - \langle AB'\rangle &\equiv & \int d\lambda \rho(\lambda)A(a,\lambda)B(b,\lambda) - \int d\lambda \rho(\lambda)A(a,\lambda)B(b',\lambda) \ ,\label{eq:identity} 
\end{eqnarray}
where  $A, B, B' = \pm1$ are the {\em results} of measurements and depend on the orientation of various filters ($a,b$ or $b'$) and may also depend  on hidden variables, $\lambda$, which are assumed to have some distribution, $\rho(\lambda)$.   Experimental averages have a similar form, for example, $\langle AB\rangle \equiv N^{-1}\sum_i A_i(a)B_i(b)$, and this provides a more compact notation.
  Bell's  derivation pulls the integral out of the difference,
  \begin{eqnarray}
  \langle AB\rangle - \langle AB'\rangle & = \langle AB -  AB'\rangle \ , \label{eq:counterfactual}
  \end{eqnarray}
  introduces extras terms that sum to zero and re-factors them,
    \begin{eqnarray}
  \langle AB\rangle - \langle AB'\rangle & = &\langle AB \pm ABA'B' \mp ABA'B' -  AB'\rangle \ , \label{eq:addzero} \\
   &=& \langle AB(1 \pm A'B')\rangle - \langle AB'(1 \pm A'B)'\rangle \ , \nonumber
  \end{eqnarray}
and takes absolute magnitude of both sides,
 \begin{eqnarray}
|\langle AB\rangle - \langle AB'\rangle| &\le &  2 \pm \langle A'B'\rangle \pm  \langle A'B\rangle \ . \nonumber
\end{eqnarray}
By taking either the $+$ or $-$ sign depending on the sign of the quantity $(\langle A'B'\rangle + \langle A'B\rangle)$,  we can write,
 \begin{eqnarray}
|\langle AB\rangle - \langle AB'\rangle| + |\langle A'B'\rangle + \langle A'B\rangle|& \le & 2 \ , \label{eq:bellimit}
\end{eqnarray}
 the famous Bell inequality.  

The act of undistributing the integral  at equation~\ref{eq:counterfactual}  required us to assume that the measurement  result $B(b,\lambda)$ is known at the same (timeless) moment that measurement result $B(b',\lambda)$ is known,  {\em the theoretical equivalent of assuming that Heisenberg's Uncertainty Principle doesn't apply} ({\it i.e.} that $\sigma_z$ and $\sigma_x$ can both be {\em measured} simultaneously, in contradistinction to knowing $\lambda_x$ and $\lambda_z$, which, torqued perhaps by the measurement apparatus,  determine how each particle moves and thus which component of  $\vec{\sigma}$ will be found to be $\pm1$).  This ``undistribution'' has also been termed counterfactual because experimentally, Bob cannot make both measurements at the same time  -- simultaneous magnetic fields at zero and ninety degrees compose a single field at 45 degrees; a polaroid filter at 0 and $90^o$ simultaneously would be opaque, {\it etc}.  

If we had considered only equation~\ref{eq:identity}, Alice and Bob could have measured $\langle AB \rangle$ here and now while Ann and Ben could have measured $\langle AB' \rangle$ in a different galaxy long ago.  But, as soon as we attempt to combine the two expressions on the right hand side of equation~\ref{eq:counterfactual}, we are, perforce, talking about making two sets of measurements on the same particle, unless we have ensured that all the hidden variables in the one ensemble of measurements match those in the other ensemble (which has to be done without doing any measurements!).  When all the macro- and micro-distinguishing characteristics have been so matched, there is, of course, no actual difference between a model with two such matched ensembles and a single ensemble of two simultaneous measurements.

\subsection{Bell's Theorem forced to  correspond with Data\label{sec:belldata}}

To avoid  Bell's  counterfactual step and respect Heisenberg's Uncertainty Principle, we will  introduce a measure of  time in our derivation by imagining that the four experiments -- $\langle AB\rangle , \langle A'B'\rangle ,  \langle AB'\rangle ,\langle A'B\rangle$ -- are measured sequentially  at times $t_1,\  t_2,\  t_3$ and $ t_4$, respectively.  We will make the same assumption of locality as Bell, that Alice's result, $A$, is only a function of her  setting, $a$, and is independent of Bob's setting, $b$ ({\it i.e.} $A = A(a,\lambda) \ne A(a,b,\lambda)$) and {\it vice versa} .  We differ from  Bell by writing  $A(a(t), \lambda(t))$ in general.  The detector settings, $a(t),\  b(t)$ will each  be constant for the periods of time corresponding to the different correlation measurements, so that  we will write $a'_2$ during the second time interval, and $b_3$ during the third time interval, {\it etc}. 

For typographical convenience, we will assume that any theoretical averages that would have been written as $ \int d\lambda \rho(\lambda)A(a,\lambda)B(b,\lambda)  $ can be converted into a normalized sum over (an arbitrarily  large number of) events, $ N^{-1}\sum_{i=1}^{N} A_i(a, \lambda_i)B_i(b,\lambda_i) $, by ensuring that $\lambda_i$ occurs with a frequency proportional to $\rho(\lambda)$.

In an earlier report~\cite{mrc3}, we began with the same asymmetric form as Bell (equation~\ref{eq:identity}), and derived an asymmetric result~\footnote{The earlier report had a multiplier, $f\ge1$, on Bob's  ``commutator'' term; our current derivation no longer needs such a factor.},
 \begin{eqnarray}
|\langle AB\rangle - \langle AB'\rangle| + |\langle A'B'\rangle + \langle A'B\rangle|& \le & 2 +    \left| \sum [B_1  B'_3 - B'_2 B_4 ] \right|  \ , \label{eq:mrc3limit}
\end{eqnarray}
but we now begin 	our derivation with the experimental/theoretical data written in a symmetric form:
\begin{eqnarray}
\langle AB\rangle_1 - \langle A'B'\rangle_2 
&=& N_1^{-1}\sum_{i_1=1}^{N_1} A(a_1,\lambda_{i_1})B(b_1,\lambda_{i_1})
 -  N_2^{-1}\sum_{i_2=1}^{N_2} A(a_2',\lambda_{i_2}) B(b'_2,\lambda_{i_2})  \ . \nonumber \end{eqnarray}
We will assume a reordering of the arbitrarily large number of  elements within the (theoretical) ensembles so that their hidden variables correspond to each other arbitrarily closely  ($\lambda_{i_1} = \lambda_{i_2} = \lambda_i $).  We  can now  complete Bell's first step,
 \begin{eqnarray}
\langle AB\rangle - \langle A'B'\rangle 
&=& N^{-1} \sum_{i=1}^{N} A(a_1,\lambda_i)B(b_1,\lambda_i) - A(a_2',\lambda_i) B(b_2',\lambda_i) 
  =  \langle AB -  A'B'\rangle  \ ,  \nonumber 
\end{eqnarray}
\noindent
allowing us to add and subtract terms. Since $\lambda_i$,  is now common in every term, 
we introduce  the further abbreviation $A(a'_n, \lambda_{i_n}) \equiv A'_n$, {\it etc.}, and absorb factors of $N^{-1}$ into the summation sign(s) so that we can compactly write,
\begin{eqnarray}
 \langle AB -  A'B'\rangle 
 &\equiv &\sum_{i=1}^{N} \left[A_1B_1  -  A_2 B'_2 \right] \ , \nonumber \\
 & = & \sum [ A_1 B_1 \pm \alpha\left( A_1 B_1 A_3 B'_3 - A_1 B_1 A_3 B'_3 \right)
                                    \pm (1-\alpha)\left( A_1 B_1 A'_4 B_4 - A_1 B_1 A'_4 B_4 \right) \nonumber \\ 
  &    & \ \ \ -  A'_2 B'_2 \mp \alpha\left( A'_2 B'_2 A'_4 B_4 -  A'_2 B'_2  A'_4 B_4 \right) 
                                     \mp (1-\alpha) \left( A'_2 B'_2 A_3 B'_3 -  A'_2 B'_2  A_3 B'_3 \right) ] \ ,\label{eq:myterms}
\end{eqnarray}
which we note has $8$ added terms instead of the $2$ in Bell's version (at equation~\ref{eq:addzero}) because our temporal subscripts have removed an ambiguity that was not apparent in Bell's counterfactual derivation (which requires at least 4 terms), and because we desire a symmetric final result (requiring the other four terms when $\alpha\ne 0,1$).  Collecting terms, 
\begin{eqnarray}
 \langle AB -  A'B'\rangle &=&          \alpha \sum A_1 B_1  [1 \pm  A_3 B'_3 ]  
                                              + (1-\alpha) \sum A_1 B_1  [1 \pm A'_4 B_4 ] \ \nonumber \\
                                      &  &       -\alpha \sum  A'_2 B'_2  [1 \pm A'_4 B_4 ] 
                                            -  (1-\alpha) \sum  A'_2 B'_2  [1 \pm A_3 B'_3 ] \  \nonumber \\
                   & &     \mp  \alpha \sum A_1 B_1 A_3 B'_3    \mp (1-\alpha) \sum  A_1 B_1 A'_4 B_4  \ \nonumber \\
                  &  &     \pm  \alpha \sum A'_2 B'_2  A'_4 B_4   \pm (1-\alpha) \sum A'_2 B'_2  A_3 B'_3    \ ,\nonumber \\
&=&  \mbox{Four Bell-ish terms }  \nonumber \\
         &  &   \     \mp \alpha       \sum [A_1 B_1 A_3 B'_3 -A'_2 B'_2  A'_4 B_4 ]  \ \nonumber \\
         &  &   \     \mp (1-\alpha) \sum [ A_1 B_1 A'_4 B_4 -A'_2 B'_2  A_3 B'_3 ]  \  . \nonumber 
\end{eqnarray}
Adding and subtracting more  terms and refactoring results in,
\begin{eqnarray}
\langle AB\rangle - \langle A'B'\rangle 
   &=&  \mbox{Four Bell-ish terms}  \nonumber \\ 
           &  &   \     \mp \alpha       \sum \left( A_1A_3 [ B_1  B'_3 - B'_2 B_4] +B'_2 B_4[ A_1 A_3 - A'_2   A'_4 ] \right) \ \nonumber \\
         &  &   \     \mp (1-\alpha) \sum \left( A_1 A'_4 [  B_1  B_4 - B'_2 B'_3] + B'_2 B'_3[ A_1 A'_4 - A'_2  A_3  ] \right)  \  .  \label{eq:nullcomms}
\end{eqnarray}

We now want to eliminate the two terms that involve differences of repeated measurements, and need to justify setting $A_iA_j= B'_k B'_l=1$; this will require us to examine  just how these other hidden variables ($\lambda_{i}(t_3), \lambda_{i}(t_4)$)  ``correspond'' to the others ($\lambda_{i}(t_1) = \lambda_{i}(t_2)$).

 We could just ignore the hidden variables and merely take the first element of the $a_3$ ensemble that has the same sign as the current element of the $a_1$ ensemble (given the same inital distribution of hidden variables, there will be the same number of positive and negative results  in each ensemble -- in the limit of very large numbers), and ``result-match''  the ensembles ({\it e.g.} $A_3(\lambda_{i}(t_3))= A_1(\lambda_{i}(t_1)), \  \lambda_{}(t_3) \mbox{ unrelated to}\  \lambda_{i}(t_1)$). 
 
  Alternatively, we could assume that the hidden variables are matched by numerical equality before going into the measurement apparatus ({\it i.e.} $\lambda_{i}(t_3) = \lambda_{i}(t_1) $), thereby insuring that $A(\lambda_{i}(t_1)) = A(\lambda_{i}(t_3)$ so that such  products will always be unity.  

Or we could, in the spirit of a ``strongly'' objective derivation~\cite{Ad00}, imagine that the hidden variable(s) going into the later measurement match the  values of the hidden variable(s) that emerged from the earlier measurement ({\it e.g.} $ \lambda_{i}(t_3=t_1 + \Delta t)\ne \lambda_{i}(t_1)$); physical considerations would then imply that  repeated measurements of the same property would give the same answer on the same particle, and if the elementary particle that emerged from the $a_1$ measurement is ``matched'' to the elementary particle going into the $a_3$ measurement, then there is {\em no} distinction between the two; it might as well have been the same particle.
    
Thus,  for any one of three reasons, $A_1 = A_3, \ A'_2=A'_4,\  B_1=B_4,\  B'_2=B'_3$, and the second and third square bracketed expressions will vanish from equation~\ref{eq:nullcomms}.  This also makes the first and fourth fore-factors unity in the remaining square-bracketed expressions, 
\begin{eqnarray}
\left| \langle AB\rangle - \langle A'B'\rangle \right |
    &=&  \mbox{Bell-ish terms}  \nonumber \\ 
               &  &   \     + \alpha     \sum  [ B_1  B'_3 - B'_2 B_4]      +  (1-\alpha)\sum  [ A_1 A'_4 - A'_2  A_3  ]   \  . \label{eq:abscomm}  
\end{eqnarray}
If we now take the absolute value of the left and right hand sides, we have
\begin{eqnarray}
\left| \langle AB\rangle - \langle A'B'\rangle \right |
    &=& 1 \pm \langle AB'\rangle +1 \pm \langle A'B\rangle  \nonumber \\ 
               &  &   \     + \alpha     \left|  \sum  [ B_1  B'_3 - B'_2 B_4]\right|       +  (1-\alpha)\left| \sum  [ A_1 A'_4 - A'_2  A_3  ] \right|  \  , \nonumber 
\end{eqnarray}
and by  picking the $+$ or $-$ sign opposite to the sign of $(\langle AB'\rangle + \langle A'B\rangle)$, we can write,
 \begin{eqnarray}
|\langle AB\rangle - \langle A'B'\rangle| + |\langle AB'\rangle + \langle A'B\rangle|
   &\le &  2   + \alpha  \left| \sum [B_1  B'_3 - B'_2 B_4 ] \right| 
                    + (1-\alpha)  \left| \sum  [A_1 A'_4 - A'_2  A_3] \right|   \ , \label{eq:gallupslimit}  
\end{eqnarray} 
or, since this must hold for all values of $\alpha$ between $0$ and $1$, 
\begin{eqnarray}
|\langle AB\rangle - \langle A'B'\rangle| + |\langle AB'\rangle + \langle A'B\rangle|
   &\le &  2   + \min \left(\left| \langle B_1  B'_3 - B'_2 B_4  \rangle\right| 
                                  ,  \left| \langle  A_1 A'_4 - A'_2  A_3\rangle \right| \right)  \ , \label{eq:gallupsqlimit}  
\end{eqnarray} 
where the two extra terms are related to whether  $a$ or $b$ is measured before or after  $a'$ or $b'$ respectively (the smaller subscript is always earlier:  $t_1<t_2 < t_3<t_4$).  We will see later that this similarity to quantum mechanical commutation relations is not accidental.
  
  The notion of ``before'' and  ``after'' would not make any sense if it only referred to random elements from two independent ({\it i.e.} {\em unmatched} weakly objective) ensembles  -- Bertie$_3$'s answer is presumably independent of Bernie$_1$'s answer, and neither has anything to do with Barney$_3$'s or Bob$_4$'s.  This means that our earlier argument to ``just match signs from the $a_1$ and $a_3$ ensembles'', while acceptable at equation~\ref{eq:nullcomms}, cannot be maintained at this point if we want a result different from Bell's.
  
   Our second argument, that the hidden variable going into the $a_3$ measurement numerically match the hidden variable going into the $a_1$ measurement, {\it etc.}, cannot be maintained here either, for it too would mean that the time labels were irrelevant, that each value of $B_1$ would always equal the value of $B_4$ while $B'_3$ would always equal $B'_2$, and the difference of two correlations would vanish term by term, again leaving us with Bell's result.
   
    The only kind of argument that has the potential to allow the ``commutator'' terms in equation~\ref{eq:gallupsqlimit} to be non-zero is the one that matches the hidden variable going into the $b'_3$ measurement to the hidden variable that came out of the $b_1$ measurement, {\it etc}.  Then and only then are the extra terms in equation~\ref{eq:gallupsqlimit} cumulating the difference between making the $a'$ ($b'$) measurement before and after the $a$ ($b$) measurement on the {\em same} element of an ensemble at {\em different} times.  To justify this theoretically, Bob's variable $\lambda_3$ has to be from a distribution, $\rho'_3(\lambda)$, where $\rho'_3$ evolved during the $b_1$ measurement from the original $\rho_1$ distribution function.  Alice's $\lambda_1$ on  the other hand, while taken from the same $\rho_1$ distribution as Bob's $\lambda_1$ ($\lambda_1^A = \pm \lambda_1^B$), has to evolve into the $\lambda_4$ that goes into the $a'_4$ measurement without intervening $a'_2$ or $a_3$ measurements -- a condition that clearly precludes a ``strongly'' objective interpretation of this proof.

To accomplish an experimental  measurement of these terms that augment Bell's original limit, one could set up Stern-Gerlach magnets or polarizing beam splitters in a tandem configuration.  This would, for example, feed photons that emerged from a detector at $b$ (or $b_{\perp}$) into follow-on PBS's set at $b'$ (and $b'_{\perp}$), and allow one to measure $\langle B'_{>} B_{<}\rangle$, {\it etc}.
 
 For classical (non-hysteretic) variables, the order of measurement is irrelevant because the measurement process is assumed to be ``non-destructive'' of that which is being measured; for (conjugate) quantum entities, it is essentially a matter of definition that their measurement results be order-dependent, since one can no longer assume that the measurement leaves the system unchanged.  Keeping the time labels in place has enabled us to see that the violation of Bell's inequality has nothing to do with non-locality and everything to do with the potentially  non-classical, Heisenberg-uncertain, non-commutative  process being measured. (Any violation of our new inequality, however,  could be attributed to nonlocality!)

\subsection{Wigner's Inequality and Instruction Sets}

Wigner~\cite{Wig76} has developed a different version of Bell's Theorem.  In his thought experiment, two detectors each have 3 different settings,   $(a,a',a''), (b,b',b'')$, and two correlated particles have their component of spin $(+/-)$ measured along the given setting's axis.  If both detectors are at the same setting they will  always measure opposite spins.  Wigner imagines that each particle  carries a set of  instructions about how to behave when encountering any possible setting, with the paired particle having opposite signs:
\begin{eqnarray}
\mbox{Alice:} & (+++) \ , \ (++-) \ , \ (+-+) \ , \ (+--) \ , \ (-++) \ , \ (-+-) \ , \ (--+) \ , \ (---)& \ , \nonumber \\
\mbox{Bob:} & (---) \ , \ (--+) \ , \ (-+-) \ , \ (-++) \ , \ (+--) \ , \ (+-+) \ , \ (++-) \ , \ (+++) & \ . \nonumber 
\end{eqnarray}
Wigner assumes that there is a probability for each of these ``instruction sets'' to occur, which he denotes by the symbol of the instruction set ({\it i.e.} $prob(+++) \equiv (+++)$).  It is then possible to say that the probability for Alice and Bob to both measure positive signs at setting $(a,b')$ is given (using Alice's probabilities) by $(+-+)+(+--)$ and  for mutual positives at setting $(a',b'')$ by $(++-)+(-+-)$. By inspection, these 4 instructions/probabilities also include the two instructions that would give mutual positive signs at the setting $(a,b'')$: $(+--)+(++-)$.  Thus Wigner's inequality is,
\begin{eqnarray}
 p_{++}(a,b') + p_{++}(a',b'') > p_{++}(a,b'') \ . \label{eq:wigners}
\end{eqnarray}
Wigner then picks the configurations $(a,a',a'') = (b,b',b'') = (0, 60^o, 120^o)$.  Since the quantum mechanical probabilities are proportional to $\sin^2(\Delta \theta /2)$, this leads to the contradiction that $\frac{1}{4} +\frac{1}{4} > \frac{3}{4}$, showing that instruction sets are incompatible with quantum mechanics.

Since we are discussing the probability of occurrence of a set giving a result, there is no obvious violation of Heisenberg's Uncertainty Principle, but it {\em does} enter,  in that to add the probabilities at equation~\ref{eq:wigners} implies that the same instruction set that gave $++$ at the $(a,b')$ setting will  {\em still} give $++$ at the $(a,b'')$ setting, even though $(a,b'')$ is measured at a later time and the particle has already gone through a $(a,b')$ or $(a',b'')$ setting.  If we make the three measurements simultaneously, then we have a counterfactual experimental situation and a {\it prima facie} violation of Heisenberg's Principle.  If the measurements are staggered in time and the staggering doesn't affect the result, then we can imagine reducing the time interval between them until the two consecutive measurements are infinitesimally close.  When the limits from above and below are equal,  Heisenberg's Principle  will be  violated in the limit.

We can also calculate Wigner's probabilities from the correlation functions via the expression
\begin{eqnarray}
p_{++}(a,b') = N^{-1} \sum_i \frac{(1+A_i(a))}{2}\frac{(1+B_i(b'))}{2} \ , \label{eq:prob2expt}
\end{eqnarray}
which, assuming that the signs occur with equal frequency at each detector ($\langle A\rangle = \langle B \rangle = 0$), means that $p_{++}(a,b') = \frac{1}{4}(1 + \langle A B' \rangle)$. Since  the left hand side of equation~\ref{eq:wigners} is bounded by unity and the individual correlations are between $\pm 1$, Wigner's inequality becomes
\begin{eqnarray}
 2 -\langle A B'' \rangle  \ge &  \langle A B' \rangle +  \langle A' B'' \rangle -  \langle A B'' \rangle & \ge -1 \ , \label{eq:wig0} \\
\Rightarrow  3 \ge &  S_W  & \ge -1 \ , \nonumber
 \end{eqnarray}
and quantum mechanical correlations now violate $3 \ge -\frac{3}{2} \ge -1$.

Recalling that $\langle A' B'\rangle \equiv -1$ for this experiment's correlation functions,  we can add this to each term in equation~\ref{eq:wig0},
\begin{eqnarray}
 1 -\langle A B'' \rangle  \ge &  \langle A B' \rangle +  \langle A' B'' \rangle  + \langle A' B'\rangle-  \langle A B'' \rangle & \ge -2 \ , \nonumber \\
\Rightarrow  2 \ge & S_{Bell} & \ge -2 \ , \nonumber
 \end{eqnarray}
generating the usual Bell inequality, while quantum mechanics gives Wigner, $S_{Bell}=-\frac{5}{2}$.

With the notational change   $(b,b') \rightarrow (b',b'')$,  equation~\ref{eq:mrc3limit} can be applied to this inequality as well,
 \begin{eqnarray}
 \left | \langle A B' \rangle_1- \langle A B'' \rangle_2 \right| + \left| \langle A' B'' \rangle_3 + \langle A' B'\rangle_4 \right|
      &\le & 2   +  \left| \langle ( B_1'   B_3''\rangle  -   \langle B_2'' B_4')  \rangle \right|  , \label{eq:mybellwig}
 \end{eqnarray}
 and to obtain Wigner's result, Bob's cross-correlation coefficients only need to differ by $\pm \half$. 
 
 We have  the same counterfactual issue here  that we saw in the original Bell Theorem:   we are assuming that we know what  three {\em measurements} ({\it e.g.} $A(a), A(a'), A(a'') =+-+$) corresponding to three mutually exclusive experimental arrangements will be when treated counterfactually in the derivation  and/or we are assuming  that these initial values won't be affected by any actual and subsequent measurements when treated in a ``factual'' manner in the derivation.  Put another way, if $a=0$ and $a'=90^o$, it is manifestly obvious that $\sigma_x$ and $\sigma_y$ have simultaneously sharp eigenvalues in Wigner's model.  If one starts by assuming that Heisenberg's Uncertainty Principle doesn't apply to the model, it should not be surprising that the model's predictions don't agree with Heisenberg's quantum mechanics or experiment.  What is surprising is that the failure  is blamed on nonlocality.

\subsection{GWZZ Squares}

Gill, Weihs, Zeilinger and Zukowski~\cite{GWZZ03} provide another derivation of Bell's Theorem.   In this variant, it is claimed that the model depends only on local realism, where by realism, they mean any model by which ``one may conceive, as a thought experiment or as part of a mathematical model, of `what the measurement outcomes would be, under any of the possible measurement settings'.  These outcomes are 8 in number: $A(a|a', b|b'), B(a|a', b|b')$, adapting their notation to ours, with ``$a|a'$'' meaning ``$a$ or $a'$".  By ``locality'' they assume that $A(a|a',b)=A(a|a',b') = A(a|a')$ and $B(b|b', a) = B(b|b', a') = B(b|b')$.  Thus there are only 4 ``locally realistic'' results, $A, A', B, B'$, which they then locate on the 4 corners of a square with ($A,A'$) diagonally across from each other, and ($B,B'$) across the other diagonal.  Given that $A$, for example, can only be $\pm 1$, one considers the number of equalities along the edges of the square -- if $A=B=A'=B'$, then it must also be the case that $B'=A$; in general there are $0,2$ or $4$ equalities, which allows them to define a statistic $\Delta = \delta_{AB} - \delta_{BA'} - \delta_{A'B'} - \delta_{B'A}$,  a sum of Kronecker deltas which can only take on values of $0$ and $-2$. The expectation value of such a statistic becomes the probability of that statistic, or 
\begin{eqnarray}
 E(\Delta) \equiv Pr(A=B) - Pr(B=A') - Pr(A'=B') - Pr(B'=A) < 0 \ , \nonumber
\end{eqnarray}
while QM predicts that for the canonical Bell parameter settings that $E(\Delta)= \sqrt{2}-1 > 0$.  
As we have seen in the case of Wigner's proof at equation~\ref{eq:prob2expt}, a probability can be converted to an expectation value:
\begin{eqnarray}
Pr(A=B) = p_{++}(a,b) + p_{--}(a,b) = \half( 1 + \langle AB \rangle ) \ , \nonumber
\end{eqnarray}
which allows us to convert GWZZ's constraint into
\begin{eqnarray}
 -1 +\half S_{Bell}  & < &  0  \ ,  \nonumber  \\
\Rightarrow   S_{Bell}  & < &  2 \  .  \nonumber
 \end{eqnarray}

It is true that ``no hidden variables appear anywhere in [their] argument beyond these eight'', but those eight hidden variables are six unmeasured measurements too many: whenever an $A$ is compared to  $B$ experimentally, the incompatible experiments $A'$ and $B'$ cannot be performed at the same time; therefore the $B$ of $Pr(A=B)$ is not the same $B$ as in $Pr(B=A')$.  If one assumes that all possible measurement results can be known (measured) simultaneously ({\it i.e.} $A,A',B,B'$), the GWZZ assumption of ``local realism'' becomes a ``classical'' realism that violates the Heisenberg Uncertainty Principle; if the results are merely assumed to be insensitive to measurement order, then the principle is violated in the limit that both measurement orders are done with vanishingly small time intervals between them.  It is one thing to assume a hidden variable like the (continuous) spin of a particle ({\it e.g.} $\vec{\lambda}$), and to specify its value(s); it is another thing entirely to assume that {\em specifying} simultaneous values for the elements of reality,  $\lambda_x, \lambda_y, \lambda_z$,  means that you can  {\em measure} their properties, $\sigma_x, \ \sigma_y$, and $\sigma_z$, simultaneously. The whole point of a hidden variable theory like de Broglie-Bohm's is that the process of making a measurement causes the $\lambda_i$ to evolve, so that what might have given $\sigma_x=+1/2$ now has equal probabilities of giving either sign	 when this measurement follows an earlier $\sigma_y$ or $\sigma_z$ measurement.

\subsection{Nonlocality without Inequalities}

Two arguments are typically made in the case of ``nonlocality without inequalities''.  The first argument, for example, Mermin's~\cite{Mermin90}, is a mixture of quantum mechanics and hidden variable properties, and  is most easily made for the case of a singlet state of three particles.  If  $|\Psi\rangle = \frac{1}{\sqrt{2}}( |+++\rangle - |---\rangle )$,   it can be shown that three different spin measurements  will have the quantum mechanical expectation value of $+1$, and a fourth will have the value of $-1$ (where $A(x) \equiv \sigma_{x}^{Alice}$, {\it etc.}):
\begin{eqnarray}
E_1(x,y,y) &=  A(x)B(y)C(y) &= +1 \ , \nonumber \\
E_2(y,x,y) &=  A(y)B(x)C(y)  &= +1 \ , \nonumber \\
E_3(y,y,x) & = A(y)B(y)C (x) & = +1 \ , \nonumber \\
E_4(x,x,x)  &= A(x)B(x)C(x) &= -1  \ , \nonumber 
\end{eqnarray}
Taking the product of these four expressions is seen to give a positive sign on the left hand side, since every $A(a|a')...$ occurs twice, while the right hand side's product is $-1$. Mindful that each expectation value represents a mutually incompatible experiment that must have been measured at a different time, it is no longer obvious that $A(a(t_i))A(a(t_j))=1$, due to intervening, potentially  randomizing, measurements. Rearranging the product in a ``factual'' manner, collecting terms, and abbreviating $A(x(t_n))$ as $A_n$ and $A(y(t_n))$ as $A'_n$, we have
 \begin{eqnarray}
LHS  = E_1\cdot E_2 \cdot E_3 \cdot E_4 = &    A_1 \cdot A'_2  A'_3 \cdot  A_4  & \cdot  
  B'_1 B_2  \cdot  B'_3 B_4 \cdot   C'_1 C'_2  \cdot C_3 C_4   \cdot  \nonumber  
\end{eqnarray}
Assuming that the product of any measurement {\em immediately} repeated with itself always gives $+1$, and making the definition
 that $\Delta_B  =  B_2 B'_3 - B'_2 B_3$, we have,
  \begin{eqnarray}
LHS =  &  A_1  A_4 
               \cdot B'_1    \cdot  ( B'_2  B_3   + \Delta_B )B_4 =     1 +  B'_1  \Delta_B B_4 \
                 ^{\genfrac{}{}{0pt}{1}{?}{=}}  
               & -1 = RHS \ . \nonumber
\end{eqnarray}
If physical effects ({\it e.g.} hysteresis) make the time-lagged $\langle x_2 y_3\rangle$ correlation different from the time-lagged $\langle y_2 x_3\rangle$ correlation, then there is no reason that $B'\Delta_B B$ might not be $-2$, allowing consistency between a {\em non-classical}, time dependent hidden variable model and the prediction of quantum mechanics (after all, the quantum relation, $\hat{\sigma}_y[\hat{\sigma}_x, \hat{\sigma}_y]\hat{\sigma}_x = 2i\hat{\sigma}_y\hat{\sigma}_z\hat{\sigma}_x = -2\hat{I}$, was the source of the original paradox).

Rather than  multiply such experiments together, one could add the four expectation values and compare, for example,  $E_1 + E_2 + E_3 - E_4$ to the quantum mechanical prediction of $ 4$.
By generalizing Bell's expression for an expectation value, $E(a,b) = \langle A(a) B(b)\rangle$, Hardy~\cite{Hardy91} has devised an inequality that holds for multiple particle experiments. Defining  $E_n(a_1,a_2,...,a_n) = \langle \Pi_{k=1}^n A_k(a_k)\rangle$, (subscripts refer to the particle identifier, not the times of the experiments in this equation) he shows that
\begin{eqnarray}
\left| E_n(a) - E_n(a')\right| \le 2 \pm E_n(a''') \pm E_n(a'') \ , \label{eq:hardyineq}
\end{eqnarray}
using exactly the same counterfactual derivation method that Bell~\cite{B71} used.
 From the analysis that led to our equation~\ref{eq:gallupsqlimit},  when Hardy's $a_k$'s  carry a time index, extra terms related to $\Delta_{A_i}$ will emerge here as well, allowing a time-dependent correlation function to agree with quantum calculations and experiment.

The second type of argument, also developed by Hardy~\cite{Hardy93}, is a two-particle variation of ``non-locality without inequalities''.  By making measurements at appropriate angles with a less than completely entangled wavefunction, it is possible to infer from a particular type of coincidence at one setting that  different coincidences should or should not have occurred (counterfactually unconditionally) at different settings, which leads to a contradiction.  Experimentally, instead of measuring a correlation function that is unity, one measures the ``other channel'' correlation which should be  zero at the same time that it is also an upper bound on a significantly non-zero probability.  This is a form of Wigner's argument, as can be most readily appreciated in the notation  used by Boschi, {\it et al.}~\cite{BBMH97}:
\begin{eqnarray}
p_{++}(a',b') \le p_{++}(a,b) + p_{+-}(a',b) + p_{-+}(a,b') \ , \nonumber
\end{eqnarray}
where the three right-hand probabilities  should be zero.   For this case, they measured the contradiction,  $.069  \pm .009 \le 0$ (moving all experimental values and error bars to one side).  As we have already seen with Wigner's inequality, Boschi's expression can be translated into a Bell inequality:
\begin{eqnarray}
\langle A'B'\rangle - \langle AB\rangle + \langle A'B\rangle + \langle AB'\rangle \le 2 \ , \nonumber
\end{eqnarray}
which becomes the experimental statement $2.136 \pm .036 \le 2$.  The factual version of this inequality, which performs incompatible measurments at distinct times, will be
equation~\ref{eq:gallupsqlimit}, and we can expect that it will not be violated.

\section{A Macroscopic Experiment}

Given that the extra term on the right hand side of our inequality, equation~\ref{eq:gallupsqlimit},  measures a discrepancy between the time-lagged cross-correlation functions formed by measuring the one angle before or after the other angle, and given the tendency to anthropomorphize such experiments by referrring to the order in which the experiments ``answer'' the ``questions'' posed by the experimenters, it is easy to imagine a macroscopic version of this experiment, in which real people answer (two) questions  from pollsters, which can be asked in one order or the other.

There is a class of questions that are ambiguous (sometimes called ``loaded'' questions), in which some people's answer to the first question changes their answer to a second question.  If the answers to both questions are already ``fixed'' in a subject's mind, then the order of questions is irrelevant (it doesn't matter if the answer is due to rote memorization or studied thought); if the answers are not present or are undecided, and the subject tries to ``ad lib'' the interview, then there is a chance of measuring such an ``EPR'' effect.

To make the experiment even more similar to the quantum ones, we could imagine taking entangled couples ({\it i.e.} ``well-married'' couples, those who can complete each others' sentences and who always answer the same questions in the same way), and let Gallup ask Alice one of the two ambiguous questions, and Harris ask Bob one of the two same questions.  It is also easy to do an ``exit interview'', and ask each interviewee the other question as well, so that we can, over a large enough sample, measure both the left and right hand sides of what we might here call ``Gallup's inequality'' (equation~\ref{eq:gallupsqlimit}).

As an example of ambiguous questions, consider the following: ``Do you favor reducing spending to balance the budget?'' and ``Do you favor increased defense spending to counter terrorism?''.  Depending on one's politics, the last news report seen, or tax return filed (``hidden variables''), a person's first answer can be considered pre-ordained.  But there will be at least some people who will then answer the second question in a manner that appears to be more consistent with their first answer, rather than appearing to  blatantly contradict it by answering as they might otherwise have, and this should lead to non-zero ``commutator'' terms on the right hand side of this ``Bell'' inequality (one shouldn't wait too long before doing the exit poll, lest each subject's state of mind ``decohere'').

We do not suggest than an experimental violation of Bell's inequality in such a polling experiment would be evidence for telepathy or other non-local effects -- it is due to the {\em local} phenomenon in each these correlated  individuals, where their consciousness can {\em change} their  answers based on the context of the questions.  Because people can ``change their mind'', their answers can exhibit a contextual - hence time - dependence.  Electrons and photons have no mind to change, but their identity or nature is such that they satisfy certain commutation relations\footnote{In a slightly different context, Leggett and Garg~\cite{LG85} proposed an experiment that would test a ``noninvasive measurability'' assumption on the macroscopic level ($10^{15}$-$10^{23}$ electrons in a SQUID detector):  that is, ``whether it is possible, in principle, to determine the state of the system with arbitrarily small perturbation on its subsequent dynamics."  Paz and Mahler~\cite{PM93} have shown that such an experiment would have to satisfy what they termed a ``temporal Bell inequality'' (with a different sense of temporal than has been used in this paper), that is bounded by 2.  When the technology exists to do such experiments and violations of the bound are observed, one hopes that the blame will be laid on the assumption that ``Heisenberg's Uncertainty Principle can be ignored'', rather than any failure of locality.}, 
 which brings us to a quantum mechanical examination of Bell's theorem.

\section{Bell's Theorem for Operators\label{secn:QM}}
 Bell's inequality, based as we have now seen on the denial of Heisenberg's Uncertainty Principle,  has been violated by all  experiments done to date. We now investigate whether there is any model -- for the quantum particles --  that explains the exact value(s) that are measured. Some authors~\cite{Malley04} have shown mathematically that if certain types of  hidden variable models describe quantum events, then the quantum observables ({\it i.e.} the corresponding operators) must commute.  More to the point, other authors ({\it e.g.}~\cite{R03, dBMR99}) have started with  a Bell ``operator'',
\begin{eqnarray}
S_{Bell} &\equiv &  \langle \psi| \hat{A}\hat{B}|\psi\rangle  
                                   + \langle \psi| \hat{A'}\hat{B} |\psi \rangle 
                                  + \langle \psi| \hat{A}\hat{B'}|\psi \rangle 
                                   - \langle \psi| \hat{A'}\hat{B'}|\psi\rangle \ , \nonumber \\
  & = &\langle \psi|\hat{A}\hat{B} + \hat{A'}\hat{B} 
                     + \hat{A}\hat{B'} - \hat{A'}\hat{B'}|\psi\rangle 
                 \equiv    \langle \psi|\hat{S}_{Bell}|\psi\rangle \ ,\label{eq:bellparmq}
\end{eqnarray}
and shown it to satisfy an operator identity,
 \begin{eqnarray}
\hat{S}_{Bell}^2 = \left( \hat{A}\hat{B}+\hat{A'}\hat{B} + \hat{A}\hat{B'} - \hat{A'}\hat{B'}\right)^2 
 & \equiv & 4\hat{I}  - [\hat{A},\hat{A'}] [\hat{B},\hat{B'}] \ , \label{eq:bellsquare}
\end{eqnarray}
on the assumption that the operators are normalized ($\hat{A}^2 = \hat{B}^2 = \hat{I}$) and local ($[\hat{A},\hat{B}] =0$).  This geometric mean expression for $S^2$ should be compared to the arithmetic mean for $S$ at  equation~\ref{eq:gallupslimit}. It has also been shown~\cite{mrc2} that  for EPR-Bell experiments, the operators that measure the projection of states along $(a, a')$, $(b, b')$ are such that 
\begin{eqnarray}
 \left[ \hat{A}, \hat{A'} \right]^{electrons} & = & 2i \sigma_{\parallel} \sin(a'-a) \ , \nonumber \\
 \left[ \hat{A}, \hat{A'} \right]^{photons} & = & 2i \sigma_{\parallel} \sin 2(a'-a) \ , \nonumber 
\end{eqnarray}
where $\sigma_{\parallel}$ is the Pauli matrix parallel to the direction of motion and the nature of the projection operator determines the argument of the sine.
When $\Delta a^{photon}=\pm 45^o$, 
equation~\ref{eq:bellsquare} yields results consistent with the experimental data:  $ \left|S_{Bell} \right| \le 2\sqrt{2}$.

Given our earlier analysis, we can now interpret this quantum mechanical result as saying that non-commuting operators are the quantum analog of ambiguous questions -- at least some  hidden variables are  likely to give different results depending on whether $a$ is measured before or after $a'$.

If $\hat{A}$ and $\hat{B}$ measure mixtures of $\sigma_z$ and $\sigma_x$ and we make the identification that the horizontally polarized state $|H\rangle$ corresponds to $|1,0\rangle$ with  $\sigma_z = +1$, and the vertically polarized state  $|V\rangle$ corresponds to $|0,1\rangle$ with $\sigma_z =-1$, then the operator $\sigma_{\parallel}$ corresponds to $\sigma_y$ and we can  evaluate the Bell matrix element for unentangled and entangled photons.

For unentangled photons, $|\psi_u\rangle \sim |H_1\rangle|V_2\rangle$, given that $\langle H\left| \sigma_y \right| H\rangle =0 = \langle V\left| \sigma_y \right| V\rangle$, we  find
\begin{eqnarray}
\langle H_1V_2 \left| S_{Bell}^2 \right| H_1V_2\rangle &=&
 4 - (2i)^2 \sin(90^o)\sin(-90^o)  \langle H_1 \left| \sigma_{y}^{(1)} \right| H_1\rangle \langle V_2 \left| \sigma_{y}^{(2)} \right| V_2\rangle   , \nonumber \\
  &=& 4 \ , \nonumber
 \end{eqnarray}
 consistent with observation~\footnote{G. Weihs, {\it priv. comm.}, June 2004}.
 
   If we take the entangled  singlet wavefunction, $|\psi_e\rangle = \frac{1}{\sqrt{2}}| (H_1V_2 - V_1 H_2)\rangle$,  then we get
 \begin{eqnarray}
\langle \psi_e \left| S_{Bell}^2 \right| \psi_e\rangle &=&
 4 -4 \cdot \frac{1}{2}
 \langle H_1V_2 - V_1 H_2\left| \sigma_y^{(1)} \sigma_y^{(2)} \right|H_1V_2 - V_1 H_2\rangle   , \nonumber \\
  & = & 8  \ \ ,\nonumber
 \end{eqnarray}  
  due to the non-zero cross-terms ($\langle H\left| \sigma_y \right| V\rangle = i$, $\langle V\left| \sigma_y \right| H\rangle = -i$).  This is also consistent with observation~\cite{WZ98}.

\subsection{Other product-form wavefunctions}

If we take circularly polarized wavefunctions, $|C_{\pm}\rangle = \frac{1}{\sqrt{2}}|H \pm i V\rangle$, we can calculate the square of the Bell parameter for an unentangled product state,  \begin{eqnarray}
\langle C^1_{\pm}C^2_{\mp}  \left| \hat{S}_{Bell}^2 \right|C^1_{\pm}C^2_{\mp} \rangle &=&
 4 -4 \cdot 
 \langle C^1_{\pm} \left| \sigma_y^{(1)} \right|C^1_{\pm} \rangle  \cdot  \langle C^2_{\mp} \left| \sigma_y^{(2)} \right|C^2_{\mp} \rangle ,\nonumber \\
  & = & 8 \  \ \ ! \nonumber
 \end{eqnarray} 
This is a surprising result,  since for this particular wavefunction, one can also show that $\langle C^1_{\pm}C^2_{\mp}  \left| \hat{S}_{Bell} \right|C^1_{\pm}C^2_{\mp} \rangle  = 0$. 
Thus, we do not expect to find that circularly polarized experiments are going to actually violate Bell's inequality. 

 However, this does provide us with reason not to be surprised that Ou and Mandel's wavefunction\cite{OU88} can be factored into a product form while describing the EPR data they measured. They used a second quantized (QED) wavefunction  to describe their data,
\begin{eqnarray}
|\psi\rangle &=& (T_x T_y)^{1/2} |1_{1x},1_{2y}\rangle
                    +(R_x R_y)^{1/2} |1_{1y},1_{2x}\rangle    -i(R_y T_x)^{1/2} |1_{1x},1_{1y}\rangle
                    +i(R_x T_y)^{1/2} |1_{2x},1_{2y}\rangle  \ , \nonumber
\end{eqnarray}
where the transmission and reflection coefficients are not assumed to be $\half$, and the one $x$- and  one $y$- polarized photon appear in various combinations at  detectors $1$ and $2$.
Although the authors did not write it so, their wavefunction could have been factored,
\begin{eqnarray}
|\psi\rangle    &\equiv&
                 ( \sqrt{T_x} | 1_{1x}\rangle + i\sqrt{R_x}|1_{2x}\rangle )\cdot
                  ( \sqrt{T_y} | 1_{2y}\rangle - i\sqrt{R_y}|1_{1y}\rangle ) 
               =  |\psi_x\rangle |\psi_y\rangle  \ , \nonumber         
\end{eqnarray}
showing that the $x$ and $y$ polarizations are not really entangled.
This wavefunction has components with both photons going to Alice (or Bob), but the coincidence measurement (or Bell-measurement operator) has zero matrix element with such components, making the entanglement a result of  ``post-selection'' of that half of all events where Alice and Bob each detected one  photon.   The original ``unselected''  wavefunction will generate local Bohmian trajectories for each particle which will violate Bell's inequality!  Thus we see that de Broglie-Bohm is non-problematic even for multiple ``entangled'' particles in this more correct context of field theory.  

Similar arguments should apply to experiments like that of Weihs, {\it et al.}\cite{WZ98}, where the entanglement is ``pre-selected'' before being fed into the optical fibers by throwing away the $\sim 99 \%$ of the photons in the non-intersecting parts of the down-conversion cones.

From this we conclude that if the wavefunction is required to describe what happens to {\em all} the photons from a source, and the coincidence logic is part of the ``measurement operator'', then the wavefunction {\em will} factorize and permit a local interpretation of all the phenomena.  If the coincidence logic is subsumed into the wavefunction, then only a part of the photon phenomena can be described, and that part will appear to be non-local.  That something {\em can} be described non-locally doesn't mean that it {\em has} to be described that way, nor is it very compelling when the non-local model is more restricted in what it can explain than the original local model.

\subsection{Wigner's Operator} 
 For the case of Wigner's instruction sets, we could calculate $| \hat{S}_W^2 | $, but the form of the constraints at equation~\ref{eq:wig0} suggests that $| (\hat{S}_W - 1)| \le 2 $.  Recalling that $\hat{A'}\hat{B'} \equiv  - I$ for operators as well as eigenvalues, similar algebra will yield
\begin{eqnarray}
\hat{\cal{S}}_{Bell}^2  \equiv ( \hat{S}_{W}- \hat{I})^2 &=& 4 + [\hat{A},\hat{A'}] [\hat{B'},\hat{B''}] \ .  \nonumber
 \end{eqnarray}  
 When the angular separations are $60^o$, we then have
\begin{eqnarray}
\hat{\cal{S}}_{Bell}^2 &=&  4 - 4 \sin^2 (60^o) \langle \psi \left |  \hat{\sigma}_y^{(1)} \hat{\sigma}_y^{(2)} \right| \psi \rangle  \ , \nonumber \\
  & = & 7 \ , \nonumber
\end{eqnarray}
which nicely bounds the quantum mechanical value, $|S_{Bell}| = \frac{5}{2}$, for the entangled state  (as well as for the unentangled product state, $|\psi\rangle=|C_{\pm}^{(1)}C_{\mp}^{(2)}\rangle$).  
It will be noted that our definition of $\hat{\cal{S}}$ results in the addition of the commutator product to 4, while the earlier analysis of the 2-position Bell experiment's $\hat{S}$ was decremented by the commutator product  (equation~\ref{eq:bellsquare}):  if either Alice or Bob (but not both) relabel their angles with the opposite sign convention, then $\hat{S}$ with one convention is the same as $\hat{\cal{S}}$ with the other.

\subsection{Hardy's Operator}

The ``less than fully entangled'' wavefunction used in the two-particle experiment of Boschi, {\it et al.}~\cite{BBMH97} takes the unnormalized form $| \psi \rangle = \alpha |HH\rangle - \beta |VV\rangle$,  with $\alpha/\beta= 0.46$.  Since  this experiment violates the standard Bell inequality, we can expect that the quantum mechanical expectation value of the Bell operator (squared),
\begin{eqnarray}
\hat{S}_{Bell}^2  = 4 -  [\hat{A},\hat{A'}] [\hat{B},\hat{B'}]  \ , \nonumber
\end{eqnarray}
will exceed 4.
Given their settings of  $a,b = 34^o$ and $a',b'= -18^o$, 
\begin{eqnarray}
{S}_{Bell}^2  &=& 4 -  (2i)^2 \sin^2 (2(52^o))
                         <\psi_e |\sigma_y^{(1)}\sigma_y^{(2)}|\psi_e> \ , \nonumber \\
                           & = & 4 + 4 \cdot 0.97^2 \cdot \frac{2\alpha \beta}{\alpha^2 + \beta^2} \ , \nonumber \\
                            & = & 6.86 = 2.62^2 \ , \nonumber
\end{eqnarray}
which  comfortably bounds from above the experimental value of ${S}_{Bell} = 2.136 \pm  0.036$. 

For the case of more than 1 set of ``off-diagonal'' elements in a Hardy ``ladder'', the operator analysis, while conceptually identical,  is significantly more complex and is not attempted here.

\subsection{GHZ  Multi-Particle States and Hardy/Mermin Inequalities}

The analog to the singlet spin state of 2 fermions is the Greenberger-Horne-Zeilinger state of three or more particles, $\Psi = \frac{1}{\sqrt{2}}(|+ \ldots +\rangle  + i |- \ldots -\rangle)$, for which Hardy~\cite{Hardy91} has defined a 3-particle parameter,
\begin{eqnarray}
\hat{S}_{Hardy} = \hat{A'}\hat{B}\hat{C} + \hat{A}\hat{B'}\hat{C} +\hat{A}\hat{B}\hat{C'} - \hat{A'}\hat{B'}\hat{C'} \ , \nonumber
\end{eqnarray}
equivalent (after relabelling $A,B,C\rightarrow A_1, A_2, A_3$) to the 3 particle version of  Mermin's~\cite{Mermin90}  general n-particle parameter, $F_n$,
\begin{eqnarray}
F_n = \int d\lambda \rho(\lambda)\frac{1}{2i}\left[ \Pi_{j=1}^n (A_j + iA'_j) - \Pi_{j=1}^n (A_j - iA'_j)\right] \ . \label{eq:mermin}
\end{eqnarray}
The algebra for three particles yields the following identity for the square of the $F$ operator~\cite{C0007006},
\begin{eqnarray}
(\hat{F}_3)^2 &= 4 I & - [A_1,A_1'][A_2,A_2'] - [A_2,A_2'] [A_3,A_3'] \nonumber \\
                                      &  &-  [A_3,A_3'][A_1,A_1']  \nonumber \\
                             & = 4 I&  - 3 \, (2i\sigma_y \sin(a_j - a'_j))^2 \le 16 I\ , \nonumber \\
\Rightarrow  |F_3| &\le &4 \equiv 2^{3-1} \ , \nonumber
\end{eqnarray}
in agreement with Mermin's quantum result, $|F_n| \le 2^{n-1}$ and violating his ``locality'' bound, $|F_3^{loc}|\le 2$, and the original  Bell-Hardy bound, equation~\ref{eq:hardyineq}.  We conclude that three particles states are just as local and non-commutative as two particle states.


\section{EPR and de Broglie-Bohm}

EPR~\cite{EPR35} showed that for entangled particles, one could infer from Alice's position measurement a counterfactual value for Bob's position measurement at the same time that Bob was actually measuring a value of momentum.  Since the Copenhagen interpretation of QM claimed that only Bob's momentum existed, EPR concluded that QM was ``incomplete''.  Bohm recast the EPR thought experiment into the language of spinning particles (or polarized photons), so that one could actually do an (EPR-B) experiment.  

Recapitulating the EPR argument in the language of EPR-B, Alice measures a component of spin with $a=0^o$ and Bob measures a component with $b=-90^o$; let us assume that $A(a)\equiv \sigma_{1z}  = +1$ and $B(b)\equiv \sigma_{2x}=+1$.  From the nature of the singlet state, EPR would infer that if Alice had counterfactually measured $\sigma_{1x}$, she would, with certainty,  have obtained the result $-1$ with certainty.  From this, EPR would conclude that an element of reality, $\lambda_x$ must exist even as Alice measures $\sigma_z$ corresponding to another element of reality, $\lambda_z$.  The existence of $\lambda_x$ and $\lambda_z$ at the same place and time does {\em not} require EPR to assume that both measurements of $\sigma_x$ and $\sigma_z$ can be {made} at the same place and time!

In the context of QED or second quantization, the  de-Broglie-Bohm interpretation makes the most sense:  the wavefunction describes a wave in an {\it aether}  of zero-point oscillators of energy $\half \hbar \omega$ while  the particle is a relatively compact region of space where the amplitude of the wavefunction is $\sqrt{\frac{3}{2} \hbar \omega}$,  and the centroid of that region of space follows the Bohmian guidance equation,
\begin{eqnarray}
\frac{d\vec{x}}{dt} & = & \frac{ \vec{j}(\psi(x))}{ \rho(\psi(x))} \ ,  \nonumber
\end{eqnarray}
where $\vec{ j}(\psi) = \frac{\hbar}{m} \mbox{Im} [(\nabla\psi^{\dag}) \psi - \psi^{\dag}(\nabla \psi )] $ and $\rho(\psi) = \psi^{\dag}\psi$.  In general, for any operator $\hat{y}$, $\hat{\dot{y}} \equiv [\hat{y}, \hat{H}]$, hence $\dot{y}(x) =  \rho(x)^{-1} { \psi^{\dag}(x) [\hat{y} , \hat{H}]\psi(x)}$.

The wave-particle duality of first-quantization is resolved into separate waves and particles in second-quantization at the same time the entangled wavefunction of first-quantization  becomes independent waves with entangled ``measurement operators'' in second-quantization.  In QED, one can think of the wavefunctions describing all the metaphysical possibilities open to the particle, while the pertinent combination of creation and destruction operators describes the epistemological possibilities of a given experiment.

Given the guidance equation above ({\it i.e.} the velocity operator, $\hat{\dot{x}}= i\frac{\hbar}{m}\nabla$), consider an isotropically decaying system, where either particle can go anywhere in $4\pi$ as long as momentum is conserved:  $\Psi_{12} = \psi_1(\vec{x}_1,\vec{k}_1)\psi_2(\vec{x}_2,\vec{k}_2) \delta (\vec{k}_1+\vec{k}_2) = 
 \psi_1(\vec{x}_1,\vec{k})\psi_2(\vec{x}_2,-\vec{k})$.  The guidance equation for particle 1 will operate on $\psi_1(x_1)$, and all $\psi_2(x_2)$  terms will factor out of numerator and denominator in such a case, leaving a local force description for particle 1.  Similarly, evolution operators for particle 2 will only operate on $\psi_2(x_2)$, and $\psi_1$ will factor out. This will give local evolution for either particle along any given direction $\vec{k}$.  If we limit our wavefunction to  two values of  $k$, for example $\vec{k} = \pm \hat{z}$, as subtended by two small, opposed detectors at the north and south poles around an emission source, then one {\em could} write a restricted wave-function as $\psi_2(x_2, +k)\psi_1(x_1, -k) \pm \psi_2(x_2, -k)\psi_1(x_1, +k) \sim |+\rangle|-\rangle \pm |-\rangle |+\rangle$, but this wavefunction wouldn't describe the events going on in the the bulk of the $4\pi$ emission solid angle.

 Thus, as long as ``pre-" or ``post-" selection of events is required to construct experiments to measure Bell parameters, the ``efficiency loophole'' that this creates will enable Bohmian mechanics to construct a completely local hidden variable interpretation of the quantum mechanical results.  As we  have noted earlier, it would be  more correct to say that the nonlocal model has a ``restricted applicability'' to only the selected events from the source, compared to the local model's unrestricted applicability to all events from the source.  
 
With respect to the spin of a particle, Bohmian mechanics gives the guidance equation for the ``hidden'' components of spin,
\begin{eqnarray}
\frac{d\lambda_j}{dt} = \frac{ \psi^{\dag} [\hat{\sigma}_j , \hat{H}]\psi}{ \psi^{\dag}\psi} \propto  i \epsilon_{jkl} B_k \lambda_l \ . \nonumber
\end{eqnarray}
This means that as Alice's particle gets into the field of her Stern-Gerlach magnet, the two unmeasured components orthogonal to the field direction will start precessing, making it meaningless to ask what ``value'' they have while the third is being measured.  While  Alice and Bob can each measure one and infer another component of spin, each pair of  components can only be considered to have been in a stationary state until one component was measured -- once Alice begins measuring her $y$ projection, the $z$ projection inferred from Bob's measurement is wiped out, and {\it vice versa} -- one knows more about the past than the present.  Whitaker's EPR argument~\cite{W04} can now be seen to be stronger than EPR's:  after Alice makes her measurement, and before Bob makes {\em any} measurement, Bob knows something that the Copenhagen interpretation says he can't know.  Once Bob's measurement is done, he only knows his measurement result, because Alice's inferred value has been wiped out.

Because the product form of the full wavefunction describes all possibilities open to the source's particles, a straight-forward Monte-Carlo approach to a Bell experiment would be very inefficient, since even  perfect detectors only subtend a small fraction of the total solid angle of the source.  Fortunately,   ``source biasing'' methods can be applied to such  computational problems, making the modelling just as efficient as that resulting from the non-local wavefunction method.

\section{Conclusions}

There were two basic elements to all proofs of non-locality:  the fact that the de Broglie-Bohm interpretation of a 2-particle singlet wavefunction generated non-local forces on each particle, and the ``local realistic'' Bell inequality, $S<2$.  We have seen that the analysis of both elements was flawed; the former by arbitrarily restricting the full wavefunction, the latter by assuming a violation of Heisenberg's Uncertainty Principle.

The implicit assumption of the temporal order-independence of   measurements  at different orientations  coupled with the explicit assumption of locality meant that Bell's claim of  a  {\em locality}  bound was actually a  {\em classicality} constraint ({\it i.e.} that one measurement has no effect on another). {\em Classical}  local hidden variable theories are precluded by experiment, but {\em non-classical} (non-commutative or quantum)  local hidden variable theories are not subject to Bell's original limit of 2, but Cirel'son's limit~\cite{C80} of $2\sqrt{2}$.   The additional terms of our inequality, equation~\ref{eq:gallupsqlimit},  or the quantum analog, equation~\ref{eq:bellsquare}, only contribute if non-classical effects occur locally; none of these inequalities requires a distant point to affect a nearby point's behavior in any way.
What is precluded by violations of Bell's inequality is not  {local realism} {\it per se}, but the Newtonian ``idealism'' of Heisenberg-violating hidden variable theories.  

 The de Broglie-Bohm interpretation of Quantum Mechanics~\cite{holland} is a Heisenberg-compliant theory, and as long as the full product-form wavefunction is used, and not some arbitrarily restricted form that incorporates all or part of the measurement operator, Bohmian mechanics will provide a {\em local} description of the EP-B data.


\bibliography{bell4}

\end{document}